\documentclass[aps,prl,graphicx,twocolumn,showpacs,preprintnumbers,amsmath,amssymb,superscriptaddress]{revtex4-1}

\usepackage{graphicx}

\begin{document}

\title{Bulk Superconductivity in the Dirac Semimetal TlSb}

\author{Yuxing Zhou}
\affiliation{Department of Physics, Zhejiang University, Hangzhou $310027$, China}
\author{Bin Li}
\affiliation{New Energy Technology Engineering Laboratory of Jiangsu Province and School of Science, Nanjing University of Posts and Telecommunications, Nanjing $210023$, China}
\author{Zhefeng Lou}
\author{Huancheng Chen}
\author{Qin Chen}
\author{Binjie Xu}
\author{Chunxiang Wu}
\affiliation{Department of Physics, Zhejiang University, Hangzhou $310027$, China}
\author{Jianhua Du}
\affiliation{Department of Applied Physics, China Jiliang University, Hangzhou $310018$, China}
\author{Jinhu Yang}
\affiliation{Department of Physics, Hangzhou Normal University, Hangzhou $310036$, China}
\author{Hangdong Wang}
\affiliation{Department of Physics, Hangzhou Normal University, Hangzhou $310036$, China}
\author{Minghu Fang}\email{Corresponding author: mhfang@zju.edu.cn}
\affiliation{Department of Physics, Zhejiang University, Hangzhou $310027$, China}
\affiliation{Collaborative Innovation Center of Advanced Microstructures, Nanjing University, Nanjing $210093$, China}

\date{\today}

\begin{abstract}
A feasible strategy to realize the Majorana fermions is searching for a simple compound with both bulk superconductivity and Dirac surface states. In this paper, we performed calculations of electronic band structure, the Fermi surface and surface states, as well as measured the resistivity, magnetization, specific heat for TlSb compound with a CsCl-type structure. The band structure calculations show that TlSb is a Dirac semimetal when spin-orbit coupling is taken into account. Meanwhile, we first found that TlSb is a type-$\rm {\uppercase\expandafter{\romannumeral2}}$ superconductor with \emph{T}$_{c}$ = 4.38 K, \emph{H}$_{c1}$(0) = 148 Oe, \emph{H}$_{c2}$(0) = 1.12 T and $\kappa_{GL}$ = 10.6, and confirmed it to be a moderately coupled \emph{s}-wave superconductor. Although we can not determine which bands near the Fermi level \emph{E}$_{F}$ to be responsible for superconductivity, its coexistence with the topological surface states implies that TlSb compound may be a simple material platform to realize the fault-tolerant quantum computations.
\end{abstract}

\maketitle

\section{\romannumeral1. INTRODUCTION}
Topological superconductors host Majorana fermions described by a real wave function, providing protection for quantum computations \cite{beenakker2016road}. So, realizing topological superconductivity (TSC) has became one of the most interesting topics in the condensed matter physics in the past decades. According to the discussion in Ref. \cite{sato2017topological}, there are intrinsic and artificial engineered topological superconductors. For the intrinsic, the topological nontrivial gap function naturally shows up. Sr$_{2}$RuO$_{4}$ \cite{maeno1994superconductivity,luke1998time,PhysRevLett.107.077003,RevModPhys.75.657} is the first proposed topological superconductor although the existence of chiral \emph{p}-wave superconductivity (SC) is still under debate. Cu$_{x}$Bi$_{2}$Se$_{3}$ \cite{PhysRevLett.104.057001}, as the first material to show SC (\emph{T}$_{c}$ $\sim$ 4 K) upon doping charge carrier into a topological insulator (TI), is a promising ground to look for two dimensional (2D) TSC due to the topological surface states surviving in TI even when carriers are doped. Many experiments, such as the conductance spectroscopy \cite{PhysRevLett.107.217001}, nuclear manetic resonance (NMR) measurements of the Knight-shift \cite{matano2016spin}, and specific heat in applied magnetic fields \cite{yonezawa2017thermodynamic}, have already given evidences for TSC seem emerging in Cu$_{x}$Bi$_{2}$Se$_{3}$. The nematic SC discovered in Cu$_{x}$Bi$_{2}$Se$_{3}$ \cite{matano2016spin,yonezawa2017thermodynamic} was also observed in the similar superconductors derived from Bi$_{2}$Se$_{3}$, such as in Sr$_{x}$Bi$_{2}$Se$_{3}$ \cite{liu2015superconductivity}, Nb$_{x}$Cu$_{2}$Se$_{3}$ \cite{qiu2015time}. Sn$_{1-x}$In$_{x}$Te \cite{PhysRevLett.109.217004} is another superconductor upon doping charge carriers into a topological crystalline insulator. In the cleanest sample (\emph{x} $\sim$ 0.04) with the lowest \emph{T}$_{c}$ (1.2 K), a pronounced zero-bias conductance peak (ZBCP) similar to that in Cu$_{x}$Bi$_{2}$Se$_{3}$ has been observed by point contact spectroscopy \cite{PhysRevLett.109.217004}. Another is the artificial engineered TSC in hybrid structures. According to the idea proposed by Fu and Kane \cite{PhysRevLett.100.096407}, if \emph{s}-wave pairing is imposed on the topological surface states of a three dimensional (3D) TI through superconducting proximity effect, the resulting superconducting state should be a 2D \emph{p}-wave SC harboring a Majorana zero mode in the vortex core. Experimentally, proximity-induced SC on the surface of 3D TIs has been studied by many groups\cite{PhysRevLett.109.056803,wang2012coexistence,PhysRevB.86.134504,wang2013fully,PhysRevX.3.021007,PhysRevX.4.041022,snelder2014josephson}. The observation \cite{wiedenmann20164pi} of 4$\pi$-period Josephson supercurrent in 3D HgTe TI is encouraging, although, it is difficult to elucidate the topological nature of the induced 2D SC.

However, there are controversies about TSC emerging and the observed ZBCP being a Majorana zero-energy mode (MZM) in the doped topological material. To realize TSC through a superconducting proximity effect in hybrid structures has many engineering challenges. A feasible way is to realize TSC in a simple compound, in which both the topological surface state and the bulk SC coexist, thus Majorana fermions emerge at the edge of superconductor. Recently, the observations of MZM in the core of vortex \cite{wang2018evidence,machida2019zero}, at the end of the atomic defect line \cite{chen2020atomic}, and near the Bi islands \cite{chen2020robust} and the interstitial Fe atoms \cite{yin2015observation} in the simplest Fe-based superconductor Fe$_{1+y}$Te$_{0.5}$Se$_{0.5}$ (\emph{T}$_{c}$ = 14 K) \cite{PhysRevB.78.224503} motivate us to search for the similar system. TlSb crystallizes in a cubic CsCl structure with space group \emph{P}\emph{m}$\bar{3}$\emph{m} (No. 221) (as shown in the inset of Fig. 1), from this structure a large number of topological semimetal/metals (TMs) were designed \cite{jin2019screening}, ranging from triple nodal points, type-$\rm {\uppercase\expandafter{\romannumeral1}}$ nodal lines, and critical type nodal lines to hybrid nodal lines. For example, CaTe is a typical type-$\rm {\uppercase\expandafter{\romannumeral1}}$ nodal line and Dirac TM \cite{du2017cate}; YIr is a typical triple-nodal-point TM \cite{jin2019screening}, YMg possesses multiple types of band crossing \cite{jin2019screening}. Therefore, we tried to grow TlSb crystals for studying its topological natures and SC, unfortunately, only polycrystalline TlSb samples were obtained.

In this paper, we performed calculations of the electronic band structure, the Fermi surface and the surface states on (001) plane, as well as measured resistivity, magnetization and specific heat for the polycrystalline TlSb sample. The band structure calculations show TlSb is a Dirac semimetal with 4-fold degenerate nodes near $\Gamma$ and R points. It is also found that TlSb is a type-$\rm {\uppercase\expandafter{\romannumeral2}}$ superconductor with the superconducting transition temperature \emph{T}$_{c}$ = 4.38 K, the lower critical field \emph{H}$_{c1}$(0) = 148 Oe, and the upper critical field \emph{H}$_{c2}$(0) = 1.12 T and the Ginzburg-Landau (GL) parameter $\kappa_{GL}$ = 10.6. The obtained specific heat jump, $\Delta$\emph{C}$_{el}$/$\gamma_{n}$\emph{T}$_{c}$ $\sim$ 1.42, indicates that TlSb is a conventional phonon-mediated superconductor with \emph{s}-wave superconducting symmetry. These results indicate that both \emph{s}-wave SC and surface states coexist in TlSb, whether the Majorana fermions emerge or not on the edges is needed to confirm in the future.

\begin{figure}
  \centering
  \includegraphics[width=8cm]{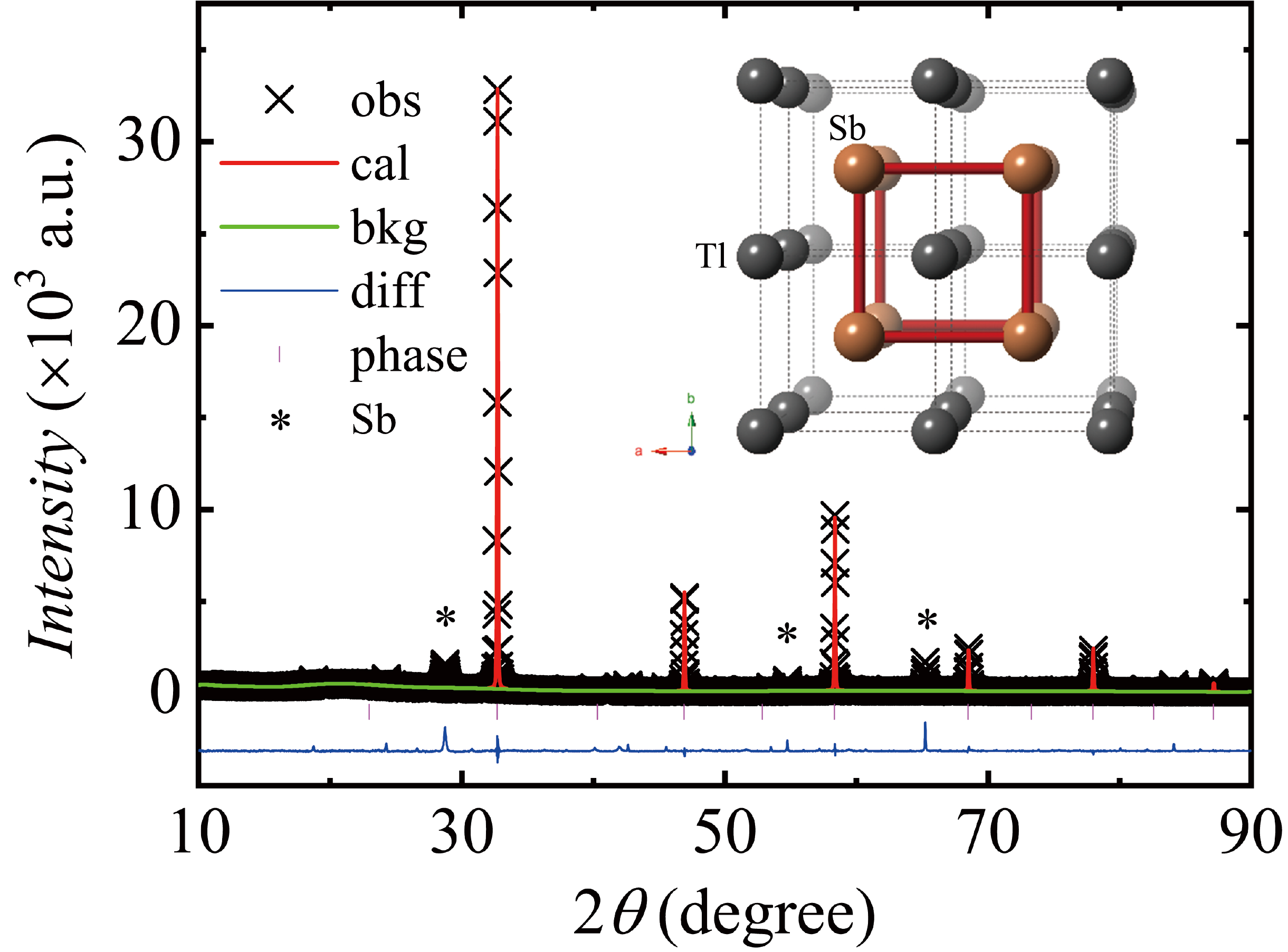}
  \caption{(a) Polycrystalline XRD pattern with its refinement profile at room temperature of TlSb. The inset shows the schematic structure of TlSb, thallium atoms are in gray while the antimony atoms are brown.}\label{Figure 1}
\end{figure}
\section{\romannumeral2. EXPERIMENTAL}
Polycrystalline TlSb samples were synthesized by a peritectic reaction method. The mixture of stoichiometric high purity Tl (99.99\%) chunk and Sb (99.999\%) powder was placed in an alumina crucible, sealed in an evacuated quartz tube and heated at 450 $^{\circ}$C for 20 hrs, then decreases to 195 $^{\circ}$C waiting for the mixture melting completely. To avoid the decomposition of obtain TlSb phase at 191 $^{\circ}$C \cite{predel1970phasengleichgewichte}, the quartz tube was quenched to room temperature at 195 $^{\circ}$C, however, the obtained TlSb samples always contain a small amount of unreacted Sb impurities due to the precipitation of Sb before the peritectic reaction. The obtained TlSb alloy is easily to cut for the subsequent structure characterizations and property measurements. Polycrystalline x-ray diffraction (XRD) was carried out on a PANalytical diffractometer equiped with CuK$_{\alpha}$ radiation. The TlSb XRD pattern is shown in Fig. 1, in which the main peaks can be fitted by the CsCl-type structure with space group \emph{P}\emph{m}$\bar{3}$\emph{m}. The lattice parameters \textit{a} = \textit{b} = \textit{c} = 3.86(5) $\rm {\AA}$ were obtained by the Rietveld refinement by using general structure analysis system (GSAS) \cite{larson1994gsas}. A rectangular bar of the sample was cut for the magnetization and resistivity mesurements, which were performed on a magnetic property measurement system (Quantum Design, MPMS - 7 T) and a physical property measurement system (Quantum Design, PPMS - 9 T), respectively. The band structure was calculated by using density function theory (DFT) with the WIEN2k package \cite{schwarz2002electronic}. Generalized gradient approximation (GGA) of Perdew-Burke-Ernzerhof (PBE) \cite{PhysRevLett.77.3865} was employed for the exchange correlation potential calculations. A cutoff energy of 520 eV and a 13 $\times$ 15 $\times$ 15 \emph{k}-point mesh were used to perform the bulk calculations. The Fermi surface (FS) was performed with WannierTools \cite{mostofi2014updated} package which is based on the maximally localized Wannier function tight-binding model \cite{PhysRevB.56.12847,PhysRevB.65.035109,marzari2012maximally} constructed by using the Wannier90 \cite{wu2018wanniertools} package.
\section{\romannumeral3. RESULTS AND DISCUSSIONS}
We first discuss the electronic band structure without considering spin-orbit coupling (SOC). As shown in Fig. 2(a), both conduction (red) and valence (blue) band cross
the Fermi level \emph{E}$_{F}$. At the high symmetry $\Gamma$ and R points, there are three bands crossing, with a threefold degenerate at 0.8 eV and 2 eV below \emph{E}$_{F}$, respectively. However, when SOC is taken into account [see Fig. 2(b)], gaps open and leaves two twofold degenerate bands at both high symmetry points. Since both time-reversal and inversion symmetries are present, no spin-splitting occurs and then the twofold degenerated bands come together to fourfold degenerate points, indicating TlSb is a Dirac semimetal. We also calculated the density of state (DOS), as shown in the right panel of Fig. 2(b), the DOS at \emph{E}$_{F}$ is mainly contributed by Sb orbits. To further clarify the band structure of TlSb, we calculated its 3D bulk FS of the first Brillouin zone (BZ) as shown in Fig. 2(d), exhibiting very complex 3D characteristics. Figure 2(e) presents the FS on the \emph{k}$_{z}$ = 0 plane, which is the cross section passing the $\Gamma$ point of the 3D FS. Figure 2(f) displays the energy dispersion in the \emph{k}$_{x}$ - \emph{k}$_{y}$ plane, in which the Dirac dispersion is clearly seen, demonstrating further that TlSb is a Dirac semimetal.
\begin{figure}
  \centering
  \includegraphics[width=8cm]{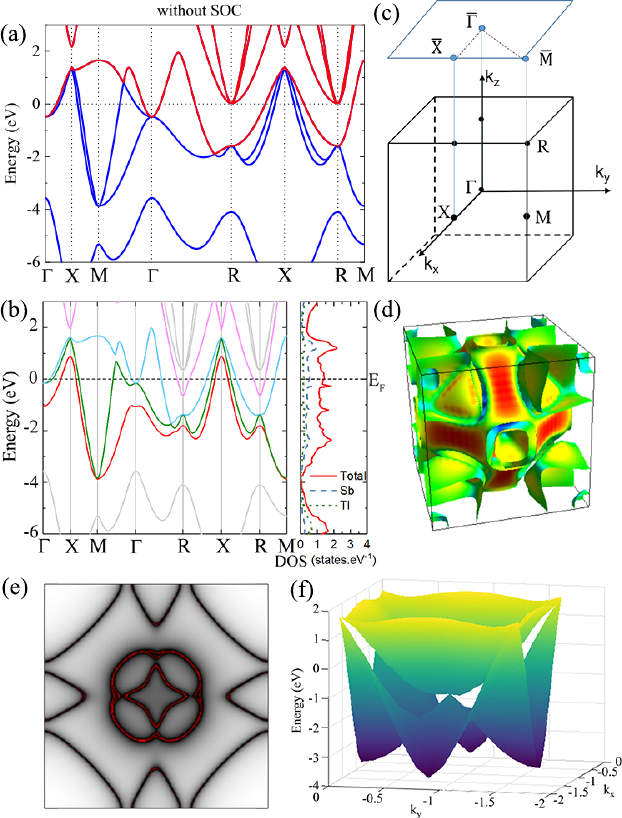}\\
  \caption{The electronic band structures of TlSb without (a) and with (b) SOC. (c) 3D bulk Fermi surfaces and color-coded Fermi velocities (red is high velocity). (d) Calculated Fermi surfaces cross section at the \emph{k}$_{z}$ = 0 plane. (e) Calculated energy distribution at Dirac point in \emph{k}$_{x}$ - \emph{k}$_{y}$ plane.}\label{Figure 2}
\end{figure}
\begin{figure}
  \centering
  \includegraphics[width=8cm]{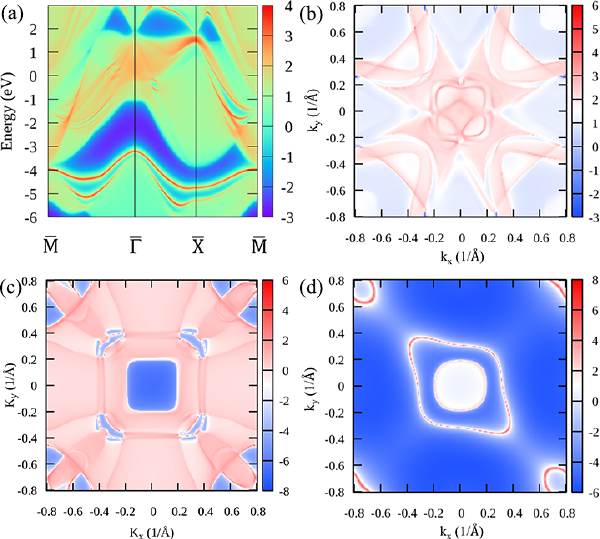}\\
  \caption{(a) Surface band structure for (001) plane along projected high symmetry points. The surface spectra of (001) plane with (b) \emph{E} = \emph{E}$_{F}$, (c) \emph{E} = -1.5 eV and (d) \emph{E} = -4 eV.}\label{Figure 3}
\end{figure}
Then we calculated the surface states on (001) plane by using a surface Green's function method \cite{sancho1985highly}. As shown in Fig. 3(a), the projected Dirac points are hidden in the continuous bulk states, the surface states are shown as the red curves. The (001) surface energy contour is shown in Fig. 3(b), (c) and (d) with \emph{E} = \emph{E}$_{F}$, \emph{E} = -1.5 eV and \emph{E} = -4 eV, respectively. The surface band at \emph{E} = -4 eV being deeply below \emph{E}$_{F}$ can be ignored due to negligible contribution to the electronic properties of material. Due to TlSb having inversion-symmetry, its topology can be described by one strong topological index $\nu_{0}$ and three weak indices $\nu_{1}$, $\nu_{2}$, $\nu_{3}$ \cite{PhysRevLett.98.106803}. Thus we calculated the Wilson loops on six time-reversal invariant planes using WannierTools \cite{wu2018wanniertools}, and the results are shown in Fig. 4. According the the definition of Wilson loops \cite{PhysRevB.84.075119,PhysRevB.83.035108}, the topological indices are (1;000) indicating that TlSb is a strong topological material.
\begin{figure}
  \centering
  \includegraphics[width=8cm]{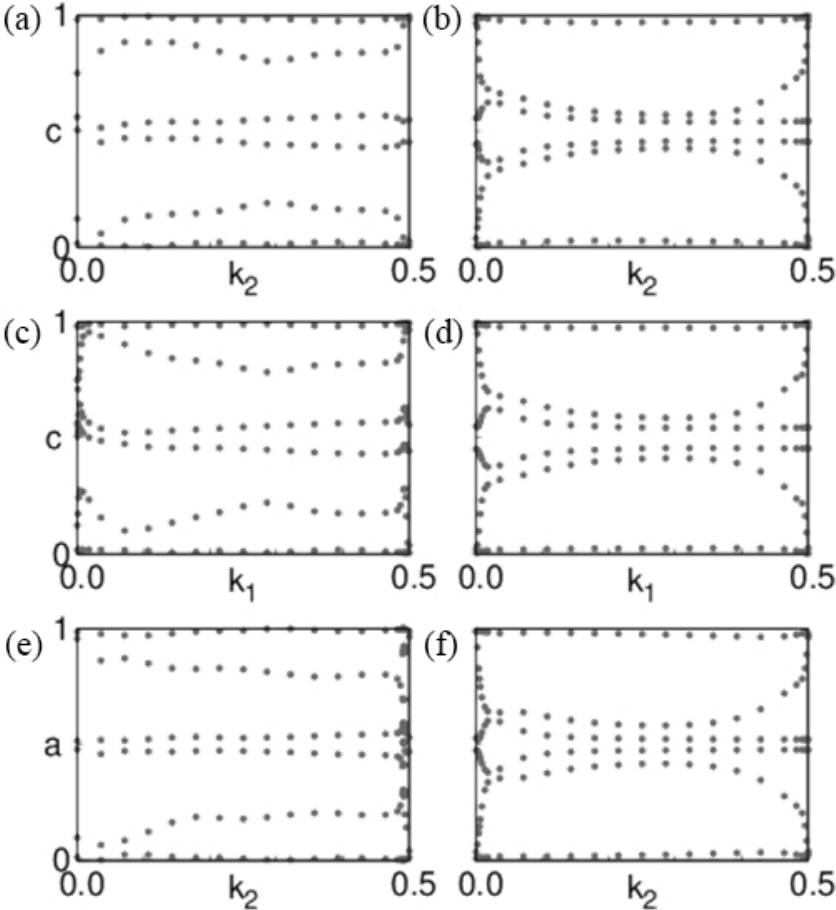}\\
  \caption{Wilson loops of six time-reversal invariant planes at (a) \emph{k}$_{1}$ = 0.0, (b) \emph{k}$_{1}$ = 0.5, (c) \emph{k}$_{2}$ = 0.0, (d) \emph{k}$_{2}$ = 0.5, (e) \emph{k}$_{3}$ = 0.0, (f) \emph{k}$_{3}$ = 0.5, where \emph{k}$_{1}$, \emph{k}$_{2}$, \emph{k}$_{3}$ are in units of the reciprocal lattice vectors.}\label{Figure 4}
\end{figure}
\begin{figure}
  \centering
  \includegraphics[width=8cm]{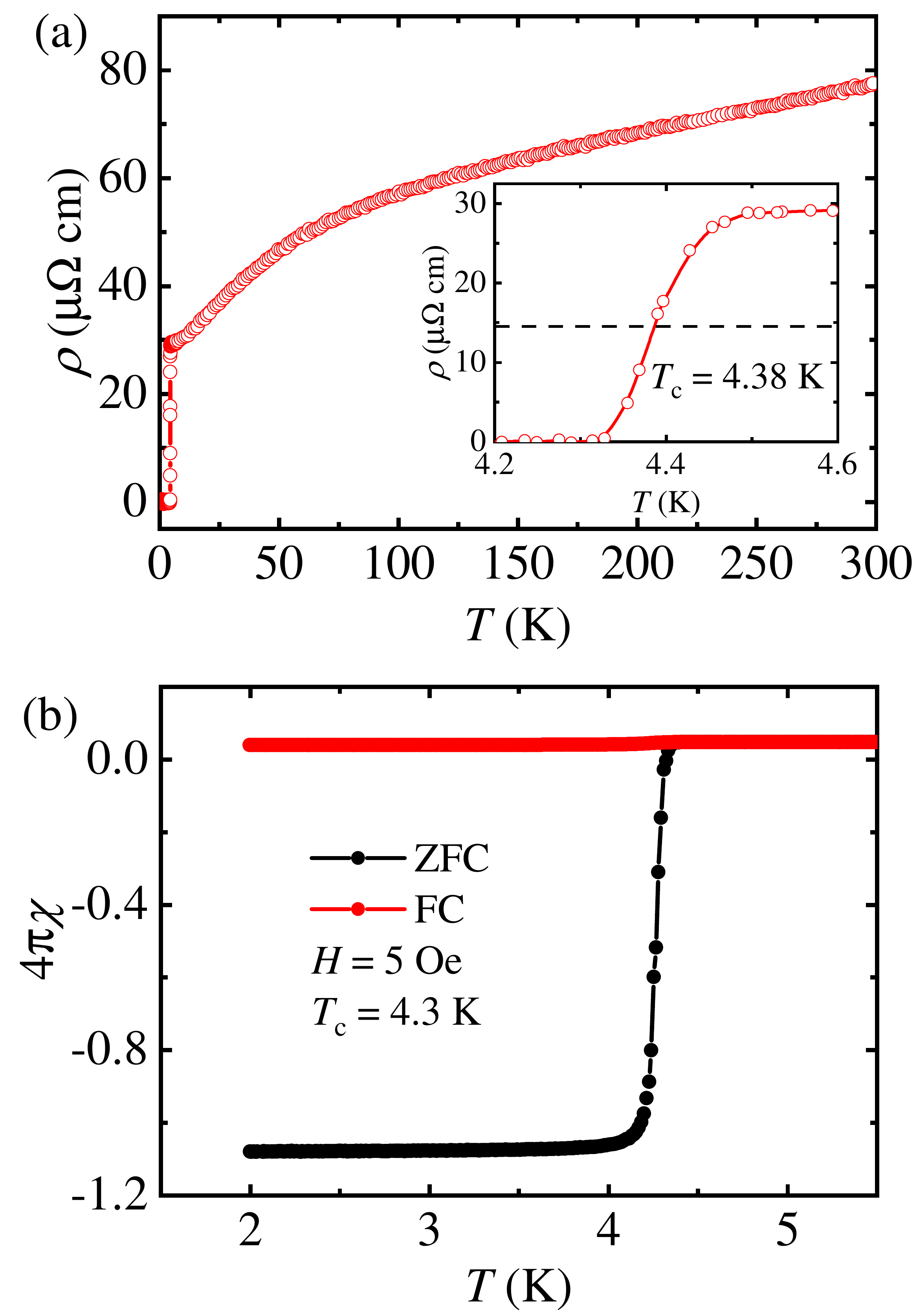}
  \caption{(a) The temperature dependence of resistivity, $\rho$(\emph{T}), of TlSb. The inset: enlarged $\rho$(\emph{T}) near the superconducting transition. (b) The temperature dependence of magnetic susceptibility, $\chi$(\emph{T}), measured at \emph{H} = 5 Oe.}\label{Figure 2}
\end{figure}

Second, we focus on the SC emerging in the Dirac semimetal TlSb discovered first by us. Figure 5(a) displays the temperature dependence of resistivity, $\rho$(\emph{T}), measured at zero field. With decreasing temperature, $\rho_{xx}$ decreases leisurely, exhibiting a poor metallic behavior, then drops to zero at 4.32 K, a superconductivity transition occurring with the mid-temperature \emph{T}$_{c}^{mid}$ = 4.38 K, $\Delta$\emph{T}$_{c}$ = 0.15 K. This superconducting transition is also confirmed by the susceptibility measurement. Figure 5(b) presents the temperature dependence of susceptibility, $\chi$(\emph{T}), measured at \emph{H} = 5 Oe with both zero-field cooling (ZFC) and field cooling (FC) process. It is clear that a  sharp diamagnetic transition emerges at 4.3 K, and the complete diamagnetism (4$\pi\chi$ $\sim$ -1) below \emph{T}$_{c}$ indicates the bulk superconductivity being from TlSb, since Sb element is non-superconducting at ambient pressure.

\begin{figure}
  \centering
  \includegraphics[width=8cm]{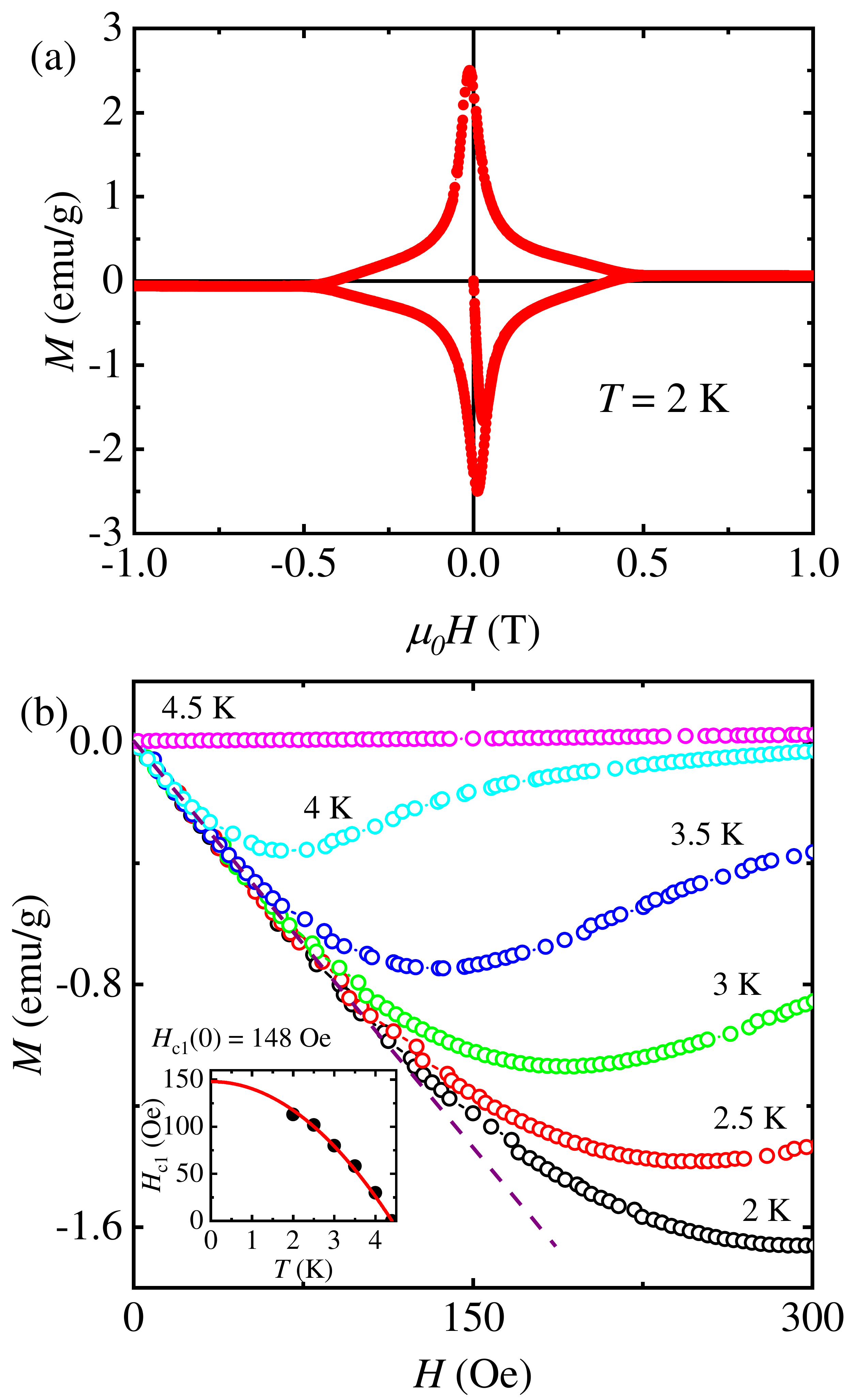}\\
  \caption{(a) Field dependence of magnetization \emph{M}(\emph{H}) measured at 2 K. (b) The low field magnetization of TlSb at different temperatures. The dashed line indicates the initial linear magnetization curve. The inset shows the temperature dependence of lower critical field, \emph{H}$_{c1}$(\emph{T}) determined by the magnetization curve deviating from linear. The red line is the \emph{H}$_{c1}$(\emph{T}) fitted by GL relation}\label{Figure 3}
\end{figure}
Figure 6(a) shows the field dependence of magnetization, \emph{M}(\emph{H}), measured at 2 K for a TlSb sample, exhibiting a hysteresis, which indicates that TlSb is a type-$\rm {\uppercase\expandafter{\romannumeral2}}$ superconductor. Then we measured the \emph{M}(\emph{H}) at various temperatures below 4.5 K, as shown in Fig. 6(b). The lower critical field \emph{H}$_{c1}$(\emph{T}) can be estimated by the field, at which \emph{M}(\emph{H}) curve starts to deviate from the linear relationship. The obtained \emph{H}$_{c1}$(\emph{T}) is shown in the inset of Fig. 6(b), the lower critical field at zero temperature \emph{H}$_{c1}$(0) = 148 Oe was obtained by the fitting using the GL relationship:
\begin{eqnarray}
  \emph{H}_{c1}(\emph{T}) =  \emph{H}_{c1}(0)(1 - (\frac{\emph{T}}{\emph{T}_{c}})^{2})
\end{eqnarray}
\begin{figure}
  \centering
  \includegraphics[width=8cm]{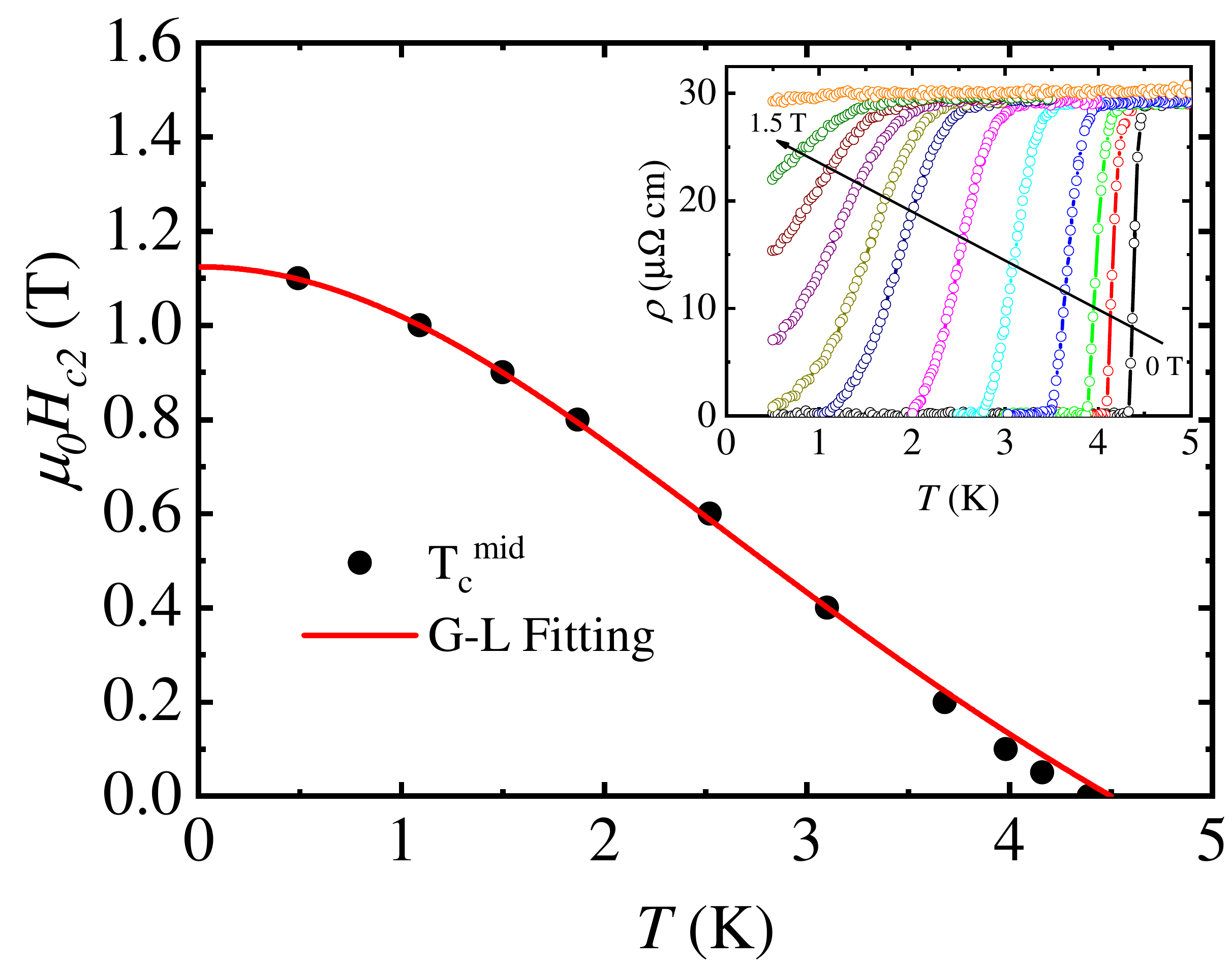}\\
  \caption{The temperature dependence of upper critical field \emph{H}$_{c2}$ determined from resistivity measurements. The inset shows the low temperature $\rho$(\emph{T}) curves measured at various magnetic fields.}\label{Figure 4}
\end{figure}

In order to get the upper critical field \emph{H}$_{c2}$(0), we measured the superconducting transition temperature (\emph{T}$_{c}^{mid}$) at various applied magnetic fields. As shown in the inset of Fig. 7, the \emph{T}$_{c}$ decreases, and the transition width $\Delta$\emph{T}$_{c}$ increases with increasing magnetic field. By using the GL formula \emph{H}$_{c2}$(\emph{T}) = \emph{H}$_{c2}$(0)(1-\emph{t}$^{2}$)/(1+\emph{t}$^{2}$), where \emph{t} is the normalized temperature \emph{t} = \emph{T}/\emph{T}$_{c}$, to fit the \emph{H}$_{c2}$(\emph{T}) data, the zero temperature upper critical field \emph{H}$_{c2}$(0) = 1.12 T was obtained, which is much lower than the Pauli limit field \emph{H}$_{c2}^{P}$(0) = 1.86\emph{T}$_{c}$ = 8.18 T. Then, the GL coherence length $\xi_{GL}$(0) = 15.3 nm was estimated by using the formula \emph{H}$_{c2}$(0) = $\Phi_{0}$/2$\pi$$\xi_{GL}^{2}$, where $\Phi$$_{0}$ is the quantum flux (\textit{h}/2\textit{e}). The penetration depth $\lambda_{GL}$(0) = 162 nm was estimated by using the formula \emph{H}$_{c1}$(0) = ($\Phi_{0}$/4$\pi\lambda_{GL}^{2}$(0))$\ln$($\lambda_{GL}$(0)/$\xi_{GL}$(0)), and the GL parameter $\kappa_{GL}$ = $\lambda_{GL}$(0)/$\xi_{GL}$(0) = 10.6.

\begin{figure}
  \centering
  \includegraphics[width=8cm]{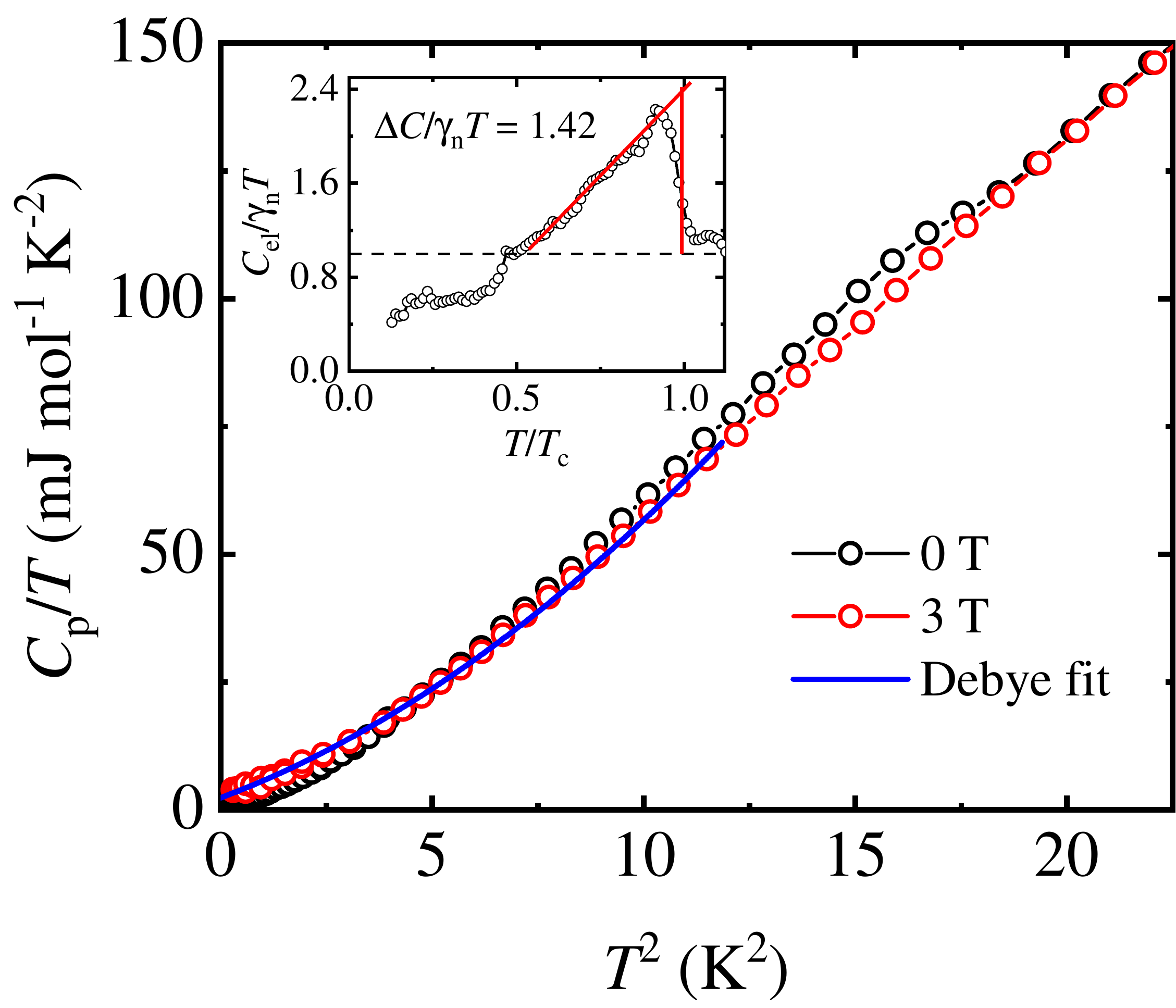}\\
  \caption{The temperature dependence of specific heat of TlSb measured at 0 T (black circles) and 3 T (red circles) plotted as \emph{C}$_{p}$ versus \emph{T}$^{2}$. The solid blue line is a fit by the Debye model. The inset shows normalized electronic specific heat \emph{C}$_{el}$/$\gamma_{n}$\emph{T} versus \emph{T}/\emph{T}$_{c}$ at zero field.}\label{Figure 5}
\end{figure}
We also measured the specific heat as a function of temperature, \emph{C}$_{p}$(\emph{T}), for TlSb in the temperature range of 0.5 - 5 K at both zero field and 3 T, respectively, as shown in Fig. 8. It is clear that the zero-field \emph{C}$_{p}$(\emph{T}), compared with the \emph{C}$_{p}$(\emph{T}) measured at 3 T ($>$ \emph{H}$_{c2}$, in this case bulk SC is completely suppressed), exhibits a small and broad peak near \emph{T}$_{c}$, corresponding to the superconducting transition. No other anomaly was observed except for the peak near \emph{T}$_{c}$ = 4.38 K, indicating that no Tl impurities (\emph{T}$_{c}$ = 2.39 K) emerge in our sample although a small amount of non-superconducting Sb impurities was detected in the XRD. We fitted the low temperature \emph{C}$_{p}$(\emph{T}) data measured at 3 T using the Debye model:
\begin{eqnarray}
  \emph{C}_{p}/\emph{T} = \gamma_{n} + \beta_{3}\emph{T}^{2} + \beta_{5}\emph{T}^{4}
\end{eqnarray}
where $\gamma_{n}$ is the Sommerfeld Coefficient, both the $\beta_{3}$\emph{T}$^{3}$ and $\beta_{5}$\emph{T}$^{5}$ are the phonon contributions to specific heat. The parameters $\gamma_{n}$ = 5.56 mJ mol$^{-1}$ K$^{-2}$, $\beta_{3}$ = 1.39 mJ mol$^{-1}$ K$^{-4}$, $\beta_{5}$ = 0.44 mJ mol$^{-1}$ K$^{-6}$ were obtained. The inset of Fig. 8 shows the normalized $\frac{\Delta\emph{C}_{p}}{\gamma_{n}\emph{T}}$ = $\frac{C_{p}(0 \rm {T}) - C_{\emph{p}}(3 \rm {T})}{\gamma_{n}\emph{T}}$ as a function of the normalized temperature \emph{t} = \emph{T}/\emph{T}$_{c}$. The bulk superconducting temperature \emph{T}$_{c}$ = 4.3 K was estimated by a entropy-balance method, consistent with the results from the resistivity and susceptibility measurements mentioned above. The normalized specific heat jump $\Delta$\emph{C}$_{el}$/$\gamma_{n}$\emph{T}$_{c}$ = 1.42 was estimated, almost the same with the predicted value (1.43) by the Bardeen-Cooper-Schrieffer (BCS) theory \cite{PhysRev.108.1175}, indicating that TlSb is a \emph{s}-wave phonon-mediated superconductor. The Debye temperature $\Theta_{D}$ = 141 K was estimated by using the formula $\Theta_{D}$ = (12$\pi^{4}$\emph{NR}/5$\beta_{3}$)$^{\frac{1}{3}}$, where \emph{N} = 2 is the number of atoms in an unit cell and the \emph{R} = 8.314 J mol$^{-1}$ K$^{-1}$ is the molar gas constant. Using the obtained $\Theta_{D}$ and \emph{T}$_{c}$ values, we calculated the electron-phonon coupling constant $\lambda_{ep}$ by using the McMillan formula \cite{PhysRev.167.331}:
\begin{eqnarray}
  \lambda_{ep} = \frac{1.04+\mu^{*}\ln(\Theta_{D}/1.45\emph{T}_{c})}{(1-0.62\mu^{*})\ln(\Theta_{D}/1.45\emph{T}_{c})-1.04}
\end{eqnarray}
where $\mu^{*}$ = 0.13 is a typical value of the Coulomb repulsion pseudopotential for the intermetallic superconductors. The obtained $\lambda_{ep}$ = 0.78 value is comparable to that of other superconductors such as PbTaSe$_{2}$ ($\lambda_{ep}$ = 0.74) \cite{PhysRevB.89.020505} and Nb$_{0.18}$Re$_{0.82}$ ($\lambda_{ep}$ = 0.73) \cite{PhysRevB.83.144525}, suggesting TlSb is a moderately coupled superconductor. The obtained superconducting parameters are summarised in Table \uppercase\expandafter{\romannumeral1}.
\begin{table}
\renewcommand\arraystretch{1.2}
  \centering
  \caption{Superconducting parameters of TlSb}\label{1}
  \setlength{\tabcolsep}{5mm}
  {
  \begin{tabular}{ccc}

    \toprule
    Parameters & unit & value\\
    \hline
    \emph{T}$_{c}$ & K & 4.3 \\
    \emph{H}$_{c1}$(0) & Oe & 148 \\
    \emph{H}$_{c2}$(0) & T & 1.12 \\
    \emph{H}$_{c2}^{P}$(0) & T & 8.18 \\
    $\xi_{GL}$ & nm & 15.3 \\
    $\lambda_{GL}$ & nm & 162 \\
    $\kappa_{GL}$ &  & 10.6 \\
    $\gamma_{n}$ & mJ mol$^{-1}$ K$^{-2}$ & 5.56 \\
    $\Delta$\emph{C}$_{el}$/$\gamma_{n}$\emph{T}$_{c}$ &  & 1.42 \\
    \botrule
   \end{tabular}
   }
\end{table}
\section{\romannumeral4. CONCLUSION}
In summary, the calculations of the electronic band structure, the FS and the surface states show that TlSb with a CsCl-type structure is a Dirac semimetal. We measured the resistivity, magnetization, specific heat for the polycrystalline TlSb sample. We first found that TlSb is a type-$\rm {\uppercase\expandafter{\romannumeral2}}$ superconductor with \emph{T}$_{c}$ = 4.38 K, \emph{H}$_{c1}$(0) = 148 Oe, \emph{H}$_{c2}$ = 1.12 T and $\kappa_{GL}$ = 10.6. The specific heat results demonstrate it to be a moderately coupled \emph{s}-wave superconductor. Although we can not determine which bands near \emph{E}$_{F}$ to be responsible for SC, the coexistence of bulk SC with \emph{s}-wave symmetry and the Dirac fermions on the surface in a single TlSb compound provides an opportunity to realize Majorana zero energy mode.

\section{ACKNOWLEDGEMENTS}
This research is supported by the National Key Program of China under Grant No. 2016YFA0300402 and the National Natural Science Foundation of China (Grants No. NSFC-12074335 and 11974095) the Fundamental Research Funds for the Central Universities, an open program from the National Lab of Solid State Microstructures of Nanjing University (Grant No. M32025).

\bibliography{citation}
\end{document}